\documentclass[showpacs,preprintnumbers,amsmath,aps,amssymb]{revtex4}
\usepackage{graphicx}% Include figure files
\usepackage[english]{babel}
\usepackage{bm}% bold math

\begin{document}
\baselineskip 16pt
\title{Cosmological super inflation using Hamilton's approach.}
\author{Omar E. N\'u\~nez}
\email{neophy@fisica.ugto.mx}
\author{J. Socorro}
\email{socorro@fisica.ugto.mx}
\author{Jos\'e Venegas}
\email{jjvenegasm88@gmail.com}
 \affiliation{Departamento de  F\'{\i}sica, DCeI, Universidad
de Guanajuato-Campus Le\'on,
 C.P. 37150, Le\'on, Guanajuato, M\'exico
}

\begin{abstract}
The Friedmann-Robertson-Walker (FRW) cosmology is analyzed with a
particular potential $\rm V(\phi)=V_0 e^{-\sqrt{3} \phi}$ in the
quintessence field scenario, which emerges in the supersymmetric
quantum mechanics (SUSY) formalism. Using Hamilton's approach for a
scalar field $\phi$ with standard kinetic energy, and the Hamilton
equations, we find exact solutions to the complete set of the
 Einstein-Klein-Gordon equations without the need of the slow-roll conditions in order to model the inflation phenomenon.
 We find that the solutions are in good agreement with the inflationary conditions such as the e-folding function ${\cal N}$
 which corresponds to the $\Omega$ function in the Misner parametrization for the scale factor $\rm A(t)=e^{\Omega(t)}$ when
 evaluated in $\rm \Delta t$, which is the time interval for the inflation period. The acceleration of the scale factor was
 computed and it was found to be positive for the inflation period with a range of values for the parameters of the model.
  Quantum solution from the Wheeler-DeWitt equation is presented, where the wave function in relation to the evolution of
  the scale factor, shows that for this period of time at larger values of $\rm A$ and for any value of scalar-field
  $\rm \varphi$, the wave function is peaked.

\end{abstract}
\pacs{4.20.Fy, 4.20.Jb, 98.80.-k, 98.80.Hw}
 \maketitle

\section{Introduction}
The inflation phenomenon is one of the most accepted mechanism to explain the early expansion of the universe, similar
to that of present day cosmic acceleration, the inflationary paradigm is considered to be a necessary part of the standard
 model of modern cosmology, solving some of its earlier crucial problems, such as the flatness or the horizon and the monopole
  ones \citep{guth}. The quintessence scalar field theory is the most commonly used in the literature to explain such phenomenon
  \citep{andrew2007, gomez, capone}. However must of them are based on the dynamical systems analysis which first step is to abandon
  the idea of finding an exact solution to the set of dynamic equations and study the behavior of the possible solutions instead,
  or are based on the dynamics of a scalar (quintessence) or multi-scalar field cosmological models of dark energy,
  (see the review \citep{copeland}), on the other hand the most recurrent approach is to use the slow-roll approximation
  whose solution does not correspond to the complete set of EKG equations.

Furthermore, we can argue that for the birth of our universe an
appropriate background is necessary, this is characterized by the
scalar field potential in the inflationary epoch, in such a way that
the evolution will be an accelerated growth. This huge growth of the
scale factor should accelerate the reheating scenario with the
presence of the radiation particles, along the inflationary process
\cite{liddle}.

Research on the inflationary topic is primarily done in two ways, one of them is to modify the General Relativity in a way that
 allows the inflationary solutions. The other way is the introduction of new forms of matter, with the capability of driving
 inflation into the General Relativity, where one introduces a canonical scalar field.
Essentially, in the studies of inflationary cosmology one imposes the usual slow roll approximation with the objective to extract
 expressions for basic observable, such as the scalar and tensor spectral indices, the running spectral index and the tensor to
  scalar ratio. The slow roll approximations reduce the set of Einstein-Klein-Gordon equations in such a way that one can quickly
   obtain the solution to the scale factor in this approximation. However there is an alternative approach which allows for an easier
derivation of many inflation results, which is called the Hamilton's formulation, widely used in analytical mechanics. Using this
 approach we obtain the exact solution of the complete set of Einstein-Klein-Gordon equations without using this approximation.

In the present work we analyze the case of scalar field cosmology, constructed using a quintessence field, and a specific potential
$\rm V(\phi)=V_0 e^{-\sqrt{3} \phi}$ in order to find exact solutions to EKG equations. There are many works in the literature
\citep{copeland,andrew1998b,ferreira2,copeland2,copeland3,quiros} that have treated this type of problems, but the employed potential
is an specific class of exponential potentials or a power law potentials, in many of those cases the evolution of the scale factor
 under this hypothesis has a time dependence that goes as a power law, and have the problem that the e-folding number is incomplete
 for the inflation scenario \cite{planckxx}, there are works in the literature that find exact solutions to a system similar to ours,
 such is the case in \cite{ratra1, ratra2} however the scalar-field potential employed in the form
  $\rm V(\phi)\propto e^{-\frac{1}{\sqrt{p}} \phi}$ is restricted to values of $ \rm p > 1$, in this format,
  our case is fixed to the value $\rm p=\frac{1}{3}$, thus the studied systems are different, as are the solutions.
  In \citep{russo} the author deals with different values of $\rm \lambda$ in a $\rm V(\phi)\propto e^{-\lambda\phi}$
  like potential as models for accelerated expansion, and in particular, the value $\rm \lambda=\sqrt{3}$ which is the
  same case we are dealing with in the present work. However, it's important to note that in his work, the author chose
   a particular transformation such that the solution he finds for the limit case $\lambda=\sqrt{3}$ is an approximated
   one, at least for this particular case he only considers a late time approximation, it is also important to note that
    the author's election of constants does not allow him to see that indeed if he were to find the exact solution to this
     particular case, must likely equivalent to ours, he would had find that it is a relevant solution for inflation and not
      as he concludes otherwise. Indeed Russo's solutions are equivalent to ours, we even find a transformation that allow us
      to represent our solutions in the same proper time as he, but a few key differences remain and thus ours exact solutions
      do allow for the inflation scenario, calculations for this case are introduced at the end of the solutions section.

By instance in \cite{liddle} the authors use the $\lambda\phi^4$ model and the e-folding number is near to 64,
however  for some cases there is a possibility to significantly increase this number. There are other works where other type of potentials
 are analyzed \cite{copeland,lobo}. In particular, in \cite{lobo} they present the solution to the Riccati equation for the Hubble function
 for various scalar potentials and the solution becomes different to that of the power law.

However, it has been shown that a potential of the form $\rm V(\phi)=V_0 e^{-\sqrt{3} \phi}$ is a viable candidate and can be argued
that it is the best suited to model the inflation phenomenon, some of those potentials come from the supersymmetric quantum mechanics
(SUSY) \cite{sodo,ssw,nuevo} or from the variable cosmological term model \cite{sdp}. In this work we do not use the slow roll approximation
to solve the Einstein-Klein-Gordon equations, and we found that the solutions are in good agreement with the inflationary conditions.

We complement our investigation  within the framework of the minisuperspace approximation of quantum theory when we analyze the models
with a finite number of degrees of freedom. Considering our cosmological model from canonical quantum cosmology under determined
 conditions in the evolution of our universe,  we obtain an exact solution to the Wheeler-DeWitt equation, where the wave function
  has a damping behavior with respect to the scale factor, which is essential for the birth of the classical universe.

This work is arranged as follows. In section \ref{model} we present the corresponding Einstein Klein Gordon equation for our
cosmological model under consideration and obtain some relations that must be satisfied by the scale factor in the evolution of
 the universe.

Then, in section \ref{ham}, we introduce the Hamiltonian apparatus which allowed us to construct a master equation, and obtain
 the solution from such equation.
The corresponding scalar potential that emerges from the temporal solution has, as is cited in the literature, an exponential behavior,
 \cite{Lucchin,Halliwell,ferreira,copeland2,Burd,Weetterich,copeland}.
In section \ref{qap} we present the corresponding quantum Wheeler-DeWitt equations, where the wave function solutions are those
that have a damping behavior with respect to the scale factor, since only such wave functions allow for good classical solutions.
Finally, we conclude in section \ref{conclusion}.

\section{The model \label{model}}
We begin with the construction of the  scalar field cosmological paradigm, which requires a canonical scalar field $\rm \phi$.
The action of a universe with the constitution of such field is,

\begin{equation}
 \rm {\cal L}[g,\phi]=\sqrt{-g}\left(R-\frac{1}{2}g^{\mu \nu}\nabla_\mu\phi\nabla_\nu\phi
 +V(\phi)\right)\label{lagra}
\end{equation}
where R is the Ricci scalar and  $\rm V(\phi)$ is the corresponding scalar field potential. The corresponding variation of
(\ref{lagra}), with respect to the metric and the scalar field gives the Einstein-Klein-Gordon field
equations
\begin{eqnarray}
&&\rm
R_{\alpha\beta}-\frac{1}{2}g_{\alpha\beta}R=-\frac{1}{2}\left(\nabla_\alpha\phi\nabla_\beta\phi-\frac{1}{2}g_{\alpha\beta}
g_{\mu\nu}\nabla_\mu\phi\nabla_\nu\phi\right)+\frac{1}{2}g_{\alpha\beta}V(\phi), \label{camrel}\\
&&\rm \square\phi-\frac{\partial V}{\partial\phi}=0.\label{klein}
\end{eqnarray}
From  (\ref{camrel}) it  can be deduced  that the energy-momentum tensor associated with the scalar field is
\begin{equation}
\rm 8 \pi G
T^{(\phi)}_{\alpha\beta}=\frac{1}{2}\left(\nabla_\alpha\phi\nabla_\beta\phi-\frac{1}{2}g_{\alpha\beta}
g_{\mu\nu}\nabla_\mu\phi\nabla_\nu\phi\right)-\frac{1}{2}g_{\alpha\beta}V(\phi)
\end{equation}\\
 The line element to be considered in this work is the flat FRW
\begin{equation}
\rm ds^2=-N(t)^2 dt^2 +e^{2\Omega(t)} \left[dr^2
+r^2(d\theta^2+sin^2\theta d\phi^2) \right], \label{frw}
\end{equation}
N is the lapse function and in a special gauge we
 can directly recover the cosmic time t, where the scale factor
$\rm A(t)=e^{\Omega(t)}$ is in the Misner's parametrization, and the scalar function has an interval, $\rm \Omega \in (-\infty,\infty)$.

\subsection{field equations}
 Making use of the metric (\ref{frw}) and a co-moving  fluid, the equations (\ref{camrel}) y (\ref{klein})
 becomes (where a dot means time derivative)
\begin{eqnarray}\rm
\frac{3\dot{\Omega}^2}{N^2}-\frac{\dot{\phi}^2}{4N^2}-\frac{1}{2} V(\phi)&=&0,\label{ein0}\\
\rm
\frac{2\ddot{\Omega}}{N^2}+3\frac{\dot{\Omega}^2}{N^2}-\frac{2\dot{\Omega}\dot{N}}{N^3}+\frac{
\dot{\phi}^2}{4N^2}
-\frac{1}{2}V(\phi)&=&0,\label{eini}\\
\rm \left[-3{\dot \Omega}\frac{{\dot \phi}}{N^2}-\frac{\ddot
\phi}{N^2}+\frac{\dot \phi}{N}\frac{\dot
N}{N^2}\right]-\frac{\partial V}{\partial \phi}&=&0 \label{kg},
\end{eqnarray}

In this work we use a particular scalar field potential $\rm V(\phi)=V_0 e^{-\sqrt{3} \phi}$, which appears in the supersymmetric
quantum mechanics as the most appropriate in order to have a super inflation with respect to the  evolution of the scale factor of
the universe \cite{sodo,ssw,nuevo}.

The algebraic structure of the EKG equations does not allow us to solve this last equations, so, in order to do so is necessary to
use another method. In the following section we will implement the Hamilton's approach to obtain
the exact solutions of these equations and we do so without the need for any approximation or ansatz be it for
the scale factor $\rm \Omega$ or the scalar field $\rm \phi$.

\section{The Lagrangian and Hamiltonian density \label{ham}}
To obtain the classical solution to Einstein-Klein-Gordon equations
(\ref{camrel}) and (\ref{klein}) we shall use the Hamilton's
approach, so we need to build the corresponding Lagrangian and
Hamiltonian densities for this cosmological model.

In this way, we use (\ref{frw}) into (\ref{lagra}) and we have
\begin{equation}
\rm {\cal L}=e^{3\Omega}\left[6\frac{{\dot
\Omega}^2}{N}-\frac{1}{2}\frac{\dot \phi^2}{N}+ N V_0
e^{-\sqrt{3}\phi}\right], \label{lagrafrw}
\end{equation}
the corresponding momenta are defined in the usual way $\rm \Pi_q=\frac{\partial {\cal L}}{\partial \dot q}$,
\begin{eqnarray}
\rm \Pi_\Omega&=& \rm 12 \frac{e^{3\Omega}}{N}\dot \Omega,
\qquad\qquad \dot
\Omega=\frac{N e^{-3\Omega}}{12}\Pi_\Omega, \nonumber\\
\rm \Pi_\phi&=&\rm -\frac{e^{3\Omega}}{N}\dot \phi,\qquad\qquad \,\,
\dot \phi=-N e^{-3\Omega} \Pi_\phi. \label{momenta}
\end{eqnarray}
and the Hamiltonian density written as $\rm {\cal L}=\Pi_q \dot
q-N{\cal H}$, when we perform the variation of this canonical
Lagrangian with respect to N, i.e. $\rm \frac{\delta {\cal
L}_{canonical}}{\delta N} =0$, it's implying the constraint ${\cal
H}=0$ which infers that the Hamiltonian density is weakly zero. So in the gauge $\rm N=24 e^{3\Omega}$ the Hamiltonian density is,
\begin{equation}
\rm {\cal H}= \Pi_\Omega^2-12
\Pi_\phi^2-24  V_0 e^{6\Omega-\sqrt{3}\phi}. \label{hamifrw}
\end{equation}
\subsection{Solutions}
Using the Hamilton's approach,
we have the following set of equations
\begin{eqnarray}
\rm \dot \Omega&=&\rm 2 \Pi_\Omega, \qquad \dot \phi= -24 \Pi_\phi, \nonumber\\
\rm \dot \Pi_\Omega&=& \rm 144 V_0 e^{6\Omega- \sqrt{3} \phi},
\qquad \dot \Pi_\phi= -24\sqrt{3} V_0 e^{6\Omega- \sqrt{3}
\phi},\label{new-variables}
\end{eqnarray}
The corresponding solutions for the set of variables $\rm (\Omega,\phi)$ and $\rm (\Pi_\Omega, \Pi_\phi)$ are
\begin{eqnarray}
\rm \Omega&=&\rm \Omega_0+ 2\sqrt{3}P_0 \Delta t + \frac{P_1}{12\sqrt{3} P_0} e^{24\sqrt{3} P_0 \Delta t}, \label{solution}\\
\rm \phi&=& \rm \phi_0-12P_0 \Delta t+ \frac{P_1}{6 P_0}  e^{24\sqrt{3} P_0 \Delta t}, \\
\rm \Pi_\Omega&=&\rm \sqrt{3} P_0 +P_1 e^{24\sqrt{3} P_0 \Delta t}, \\
\rm \Pi_\phi&=&\rm -\frac{\sqrt{3}}{6} \Pi_\Omega + P_0.
\end{eqnarray}
where $\rm \Omega_0, \phi_0 $ , $\rm P_0$ and $\rm P_1$ are  integration constants. In particular, the constant $\rm \Omega_0<0$,
in that sense we obtain an appropriate initial value for the scale factor $\rm A_0=e^{\Omega_0}>0$, and
the constant $\rm \phi_0 \geq 12P_0 \Delta t$ so that we avoid any ghost field (negative scalar field values), from such conditions
 we can see that $\rm \Delta\Omega \propto \frac{\sqrt{3}}{6}\Delta\phi$.
The last solutions were introduced in the Einstein field equation (\ref{ein0},\ref{eini},\ref{kg}),
in order for such equations to hold true then $\rm V_0=\frac{\sqrt{3}}{6}P_0 P_1 e^{-6\Omega_0+\sqrt{3}\phi_0}$, and considering that
we employed a positive scalar potential, the constants $\rm P_0$ and $\rm P_1$  must be positive, so, the scale factor becomes
$$\rm A(t)=A_0  Exp\left[ \frac{P_1}{12\sqrt{3}P_0}e^{24\sqrt{3} P_0 \Delta t}+2\sqrt{3}P_0\,\Delta t\right]\, .$$

In \cite{russo} the author deals with a system similar to ours, and
using a particular transformation, one we call \emph{transformation
gauge} since is similar to ours \emph{Hamilton's gauge}, he finds
the following set of solutions, $\rm
A(\tau)=e^{\frac{1}{3}\tau^2}(2\tau^{\frac{1}{6}})$ and $\rm
\phi(\tau)=\frac{1}{\sqrt{3}}(2\tau^2-\ln \tau)$, we can see that
under a particular transformation we can change our solutions to his
proper time $\tau$, which is $\rm \tau=e^{12\sqrt{3}P_0 t}$, with
this transformation our solutions are
$$ \rm A(\tau)=e^{\frac{\sqrt{3}P_1}{36P_0}\tau^2}(2\tau^{\frac{1}{6}})\, ,
 \qquad \rm \phi(\tau)=\frac{1}{\sqrt{3}}(2\tau^2-\ln \tau)\, , $$
which are indeed equivalent but key differences remain, in particular the exponential constant in the scale factor,
the author's being simply $1/3$, and the differences are mostly because of the author's chose of his own particular
constants values and the approximation for late times for this case, in his paper the author loses focus in this particular
case with $\rm \lambda=\sqrt{3}$ and concludes that none of the solutions he found are suited for inflation, but our exact
solutions do allow for the inflation scenario.

\subsection{Inflationary conditions}
In order to prove that the solutions are inflationary one must check that the second time derivative of the scale factor remains
positive during such epoch, in that sense $d^2 A/dt_{phys}^2>o$ where $t_{phys}$ represents the physical time, let us recall that
our time is related to the physical one by $dt_{phys}=Ndt $, with $N=24e^{3\Omega}$.  We can express the scale factor acceleration
 in our time by computing the following derivative
\begin{equation}
\rm \frac{d^2 A(t_{phys})}{dt_{phys}^2}=\frac{d}{Ndt}\left(\frac{dA}{Ndt}\right),
\end{equation}

where $A(t)=e^{\Omega}$ since we are using the Misner's parametrization, so computing the time derivatives we arrive to the following
 relation for the acceleration
\begin{equation}
\rm \frac{d^2 A(t_{phys})}{dt_{phys}^2}=\frac{e^{-5\Omega}}{24^2}\left(\ddot{\Omega}-2\dot{\Omega}^2\right),
\end{equation}
introducing the $\Omega$ solution from eq.(\ref{solution}) we arrive to the following relation, which must be greater than zero
\begin{equation}
\rm 32\sqrt{3}P_0 P_1 e^{24\sqrt{3}P_0 \Delta t}-8P_1^2 e^{48\sqrt{3}P_0 \Delta t}-24P_0^2>0\, .
\end{equation}
In order to check such relation in terms of the parameters $\rm
(P_0,P_1)$ we solve the quadratic equation, using the hypothesis
that $\rm 24\sqrt{3}P_0 \Delta t=\alpha$, with $\alpha$ is a
constant, which we shall use next for obtain the e-folding number,
obtaining that
\begin{equation}
\rm \frac{P_0}{P_1}>(\sqrt{2}\mp 1)\frac{\sqrt{6}}{3}e^\alpha,
\label{condition-one}\end{equation} another strong inflationary
condition is that $\rm |dH/dt_{phys}|<<H^2$, where $\rm
H=\frac{dLnA}{Ndt}=\frac{\dot{\Omega}}{N}$, using the last
expression and the solution for $\rm \Omega$ in eq.(\ref{solution})
we arrive to the following relations
\begin{eqnarray}
\rm \dot{H} &=& \rm \frac{\ddot{\Omega}}{N^2}-\frac{\dot{\Omega}}{N^3}\dot{N}=\frac{1}{N^2}(24\sqrt{3}P_0 P_1 e^{24\sqrt{3}P_0 \Delta t}
-36P_0^2),\\
\rm H^2 &=& \rm \left(\frac{\dot{\Omega}}{N}\right)^2=\frac{1}{N^2}(8\sqrt{3}P_0 P_1 e^{24\sqrt{3}P_0 \Delta t}+12P_0^2),
\end{eqnarray}
obtaining
\begin{equation}
\rm P_0>>0, \qquad \qquad or \qquad \qquad
\frac{P_0}{P_1}>>\frac{\sqrt{3}}{3}e^\alpha, \label{condition-two}
\end{equation}
the condition (\ref{condition-one}) is contained in the this last
conditions, which can be propose a simple solution, which is that
$\rm P_0>>1$ and $\rm P_1<<1$ so that $\rm P_1/P_0 \approx 0$, $\rm
P_1\,P_0\approx constant$,  and $\rm P_1^2 \approx 0$, from here we
can infer that the acceleration will be positive only after certain
length of the inflationary epoch ($\Delta t$), but will remain so
after, in this sense the required time interval should be equal or
larger than the value that the aforementioned condition imposes.

The e-folding function ${\cal N}=\int H(t_{phys}) dt$ is equivalent
to the $\rm \Delta \Omega$ function when evaluated with $\Delta t$,
with $\rm H(t_{phys})=\frac{\dot \Omega}{N}$, which is
\begin{equation}
\rm {\cal N}= 2\sqrt{3}P_0 \Delta t + \frac{P_1}{12\sqrt{3} P_0} e^{24\sqrt{3} P_0 \Delta t}, \label{e-folding}
\end{equation}
where $\Delta t$ is the interval in arbitrary units where the inflation scenario of the universe occurs. The same conditions for the
 parameters $\rm (P_0,P_1)$ can be imposed, so that we can have the appropriate e-folding number, one that is in
 agreement with the cosmological data \cite{planckxx}. In this sense we only need to fix the value of $\rm P_0\Delta t \approx 10\sqrt{3}$,
 thus $\alpha=720$,  and depending on this last relation we can have the appropriated range
$$\rm {\cal N} \approx 2\sqrt{3}P_0 \Delta t \in (54, 60)\, .$$
Now it only remains to fix the range for the time period $\rm \Delta t$, and we can achieve this by using the inflationary period in
 physical time ($\rm \Delta t_{phys}$) and the Hubble parameter in the relation $\rm \Delta t_{phys}=1/H$, so, imposing the same conditions
 over the parameters $\rm (P_0,P_1)$ and the solution for $\rm \Omega$ in eq.(\ref{solution}) we arrive to the following relation
\begin{equation}
\Omega_0=\frac{1}{3}\ln{(2\sqrt{3}P_0 \Delta t_{phys})}-60\, ,
\label{om0}
\end{equation}
however, from supersymmetric quantum mechanics (SUSY) as discussed in \citep{obre} we can fix the value of $\rm \Omega_0 \approx -41.45$,
 and from here we can use eq.(\ref{om0}) to find the relation
$P_0\approx\frac{e^{55.65}}{2\sqrt{3}\Delta t_{phys}}\, ,$ from the
computed e-folding number we fixed the value $\rm P_0\Delta t
\approx 10\sqrt{3}$ so, using the last relation for $\rm P_0$ we
 can relate the time range of inflation with the physical one by
$$\Delta t\approx 4\times 10^{-23}\Delta t_{phys}\, ,$$ and using
 $\rm \Delta t_{phys} \approx 10^{-34} s$, we can see that our
  $\rm \Delta t$ range, in natural units, should be around $\rm \Delta t\approx6.2\times 10^{-31}$, which is rather small, however,
  is in good agreement with inflationary conditions.

\section{quantum approach\label{qap}}
The Wheeler-DeWitt (WDW) equation has been treated in many different ways and there are a lot of papers that deal with different
approaches to solve it, for example in \cite{Gibbons}, they asked the question of what a typical wave function for the universe is.
In Ref. \cite{Zhi} there appears an excellent summary of a paper on quantum cosmology where the problem of how the universe emerged
from big bang singularity can no longer be neglected in the GUT epoch. On the other hand, the best candidates for quantum solutions
 are those that have a damping behavior with respect to the scale factor, since only such wave functions allow for good classical
 solutions when using the WKB approximation for any scenario in the evolution of our universe \cite{HH,H}.

The Wheeler-DeWitt equation for this model is acquired by replacing  $\rm \Pi_{q^\mu}=-i\hbar \partial_{q^\mu}$
in (\ref {hamifrw}).  The factor $\rm e^{-3\Omega}$ may be factor ordered with $\rm \hat \Pi_\Omega$ in many ways.
Hartle and Hawking \citep{HH} have suggested what might be called a semi-general factor ordering, which
in this case would order $\rm e^{-3\Omega} \hat \Pi^2_\Omega$ as
\begin{eqnarray}
\rm - e^{-(3- Q)\Omega}\, \partial_\Omega e^{-Q\Omega} \partial_\Omega&=&\rm - e^{-3\Omega}\, \partial^2_\Omega +
 Q\, e^{-3\Omega} \partial_\Omega, \label {hh}
\end{eqnarray}
where  Q is any real constant that measure the ambiguity in the factor ordering for the variable $\Omega$.
In the following we will assume such factor ordering for the Wheeler-DeWitt equation, which becomes
\begin{equation}
\rm \hbar^2 \Box \Psi+ \hbar^2 Q\frac{\partial \Psi}{\partial \Omega}- e^{6\Omega}U(\varphi)\Psi=0, \label{wdwmod}
\end{equation}
where the field was re-scaled as $\rm \phi=\sqrt{12}\varphi$, $\rm
\Box=-\frac{\partial^2}{\partial
\Omega^2}+\frac{\partial^2}{\partial \varphi^2}$ is the
d'Alambertian in the coordinates $q^\mu=(\Omega,\varphi)$ and the
potential is $\rm U=  +24V_0 e^{-6\varphi} $.

Making the canonical transformation
\begin{equation}
\rm x=6\Omega-6\varphi, \qquad y=\alpha_1 \Omega + \alpha_2 \varphi,
\label{can-trans}
\end{equation}
eq. (\ref{wdwmod}) is rewritten as
$$\rm - \hbar^2 \alpha \frac{\partial^2}{\partial x \partial y}\Psi(x,y) -\beta \hbar^2 \frac{\partial^2}{\partial y^2}\Psi(x,y)
+\hbar^2 Q\left(6\frac{\partial}{\partial x}+
\alpha_1\frac{\partial}{\partial y} \right)\Psi(x,y)-24 V_0
e^{x}\Psi=0,$$ where the constants  $\alpha=\alpha_1+\alpha_2$ and
$\beta=\alpha_1^2-\alpha_2^2$. This equation have the following
solution $\rm \Psi(x,y)=e^y \, e^{u(x)}$, where the function $\rm
u(x)$ satisfy the ordinary
 differential equation
$\rm -\hbar^2(\alpha - Q)\frac{du}{dx}-\hbar^2 \left(\beta - Q \alpha_1 +\frac{24 V_0}{\hbar^2}e^x\right)=0$, which has the
 following solution
\begin{equation}
\rm u(x)=u_0 -\frac{1}{\alpha - Q}\left[\left(\beta -Q \alpha_1\right)x + \frac{24 V_0}{\hbar^2}e^x \right],
\end{equation}
so, the wave function becomes
$$\rm \Psi(x,y)=\Psi_0 e^y Exp\left(-\frac{1}{\alpha - Q}\left[\left(\beta -Q \alpha_1\right)x + \frac{24 V_0}{\hbar^2}e^x \right]\right),$$
that written in the original variables $(\Omega,\varphi)$ is
$$\rm \Psi(\Omega,\varphi)=\Psi_0 Exp\left[a_1\Omega + a_2 \varphi \right] Exp\left[- \frac{24 V_0}{\hbar^2{(\alpha - Q)}}
e^{6\Omega-6\varphi} \right],$$
where the constants are, $\rm a_1=\frac{\alpha_1(\alpha+5Q)-6\beta}{\alpha-Q}$ and
$\rm a_2=\frac{\alpha_2 \alpha- Q(\alpha_2+6\alpha_1)+6\beta}{\alpha -Q}$.

If we take into account that the best candidates for quantum solutions are those that have a damping behavior with respect to
the scale factor, the constant must satisfy the constraint $\alpha-Q >0$, the wave function in the evolution of the scale factor
 $\rm A=e^\Omega$, shows that for this period of time at larger values of $\rm A$ and any value of scalar-field $\rm \varphi$,
 the wave function is peaked, as seen in figure 1, which is essential to give way for the classical universe. If we
 compute $\rm \frac{\partial \Psi(\Omega,\varphi)}{\partial \Omega}=0$, we can find the peak of the wave function with respect
 to $\rm \Omega$ for any value of $\rm \varphi$, so we find the following relation
$$\rm A=\left(\frac{a_1 \hbar^2}{144V_0}(\alpha-Q)\right)^{1/6}e^{\varphi}\, ,$$
where $\rm \varphi=\frac{\sqrt{3}}{6}\phi$, in figure 2 we can see
that for fixed values of the parameters the wave
 function is peaked along the classical trajectory for any values of $\rm \varphi$ which has an exponential dependence with the
 wave function.

\begin{figure}[h]
\begin {center}
\includegraphics[totalheight=0.3\textheight]{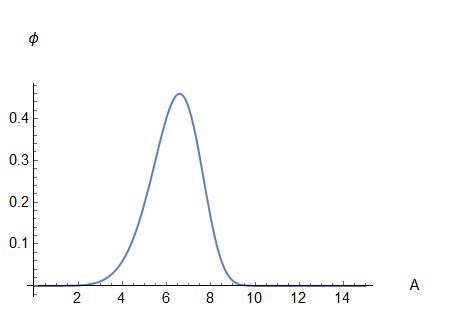}
\caption{For fixed values of the parameters, using the relation $\rm
A=e^{\Omega}$ and for a fixed value of the scalar field $\rm \phi$,
we can see that for this period of time at larger values of $\rm A$,
the wave function is peaked.}
\end{center}
\end{figure}

\begin{figure}[h]
\begin {center}
\includegraphics[totalheight=0.3\textheight]{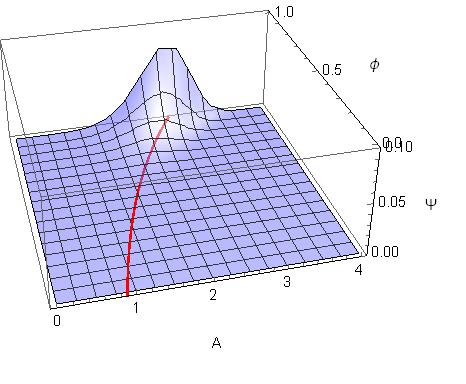}
\caption{We can see that for fixed values of the parameters the wave
function is peaked along the classical trajectory (red line in the
$\rm(A,\varphi)$ plane) for any values of $\rm \varphi$ which has an
exponential dependence with the wave function.}
\end{center}
\end{figure}

\section{Conclusions \label{conclusion}}

The inflationary paradigm has been studied under the FRW cosmological model with the aid of a particular potential from SUSY \cite{sodo}
 and the quintessence scalar field associated with it, exact solutions for the EKG set of equations were found using the Hamilton's
 approach without using the slow-roll approximation. The obtained solutions are in good agreement with the inflationary conditions such
  as the e-folding function ${\cal N}$  which is equivalent to the $\Delta\Omega(\Delta t)$ function and it was shown that for particular
   values of the parameters the appropriate e-folding number can be acquired, one that is in agreement with the cosmological data
    \cite{planckxx}, other inflationary conditions are also tested and positively checked, and with the help of SUSY we where able
    to fix the time range of inflation, in that sense many of the results were derived from the help of SUSY and for that reason we
    have what we call a \emph{Super inflation}. The quantum solution from the Wheeler-DeWitt equation was obtained and the wave function
     in relation to the evolution of the scale factor, shows that for this period of time at larger values of $\rm A$ and any value
     of scalar-field $\rm \varphi$, the wave function is peaked, which is essential to give way for the classical universe, it was
     also shown that the wave function is peaked along the classical trajectory for any values of the scalar-field.

 \acknowledgments{ \noindent
 This work was partially
supported by CONACYT  167335, 179881 grants. PROMEP grants UGTO-CA-3
. This work is part of the collaboration within the Instituto
Avanzado de Cosmolog\'{\i}a and Red PROMEP: Gravitation and
Mathematical Physics under project {\it Quantum aspects of gravity
in cosmological models, phenomenology and geometry of space-time}.
Many calculations where done by Symbolic Program REDUCE 3.8. and Wolfram Mathematica 10.0}

\end{document}